\documentclass[preprint,12pt]{emulateapj}

\usepackage{latexsym,amssymb}
\usepackage{amsmath}
\usepackage{natbib}
\usepackage[colorlinks=true,linkcolor=blue,citecolor=blue]{hyperref}

\shorttitle{Transport and Mixing of  \lowercase{{\it r}}-process elements in neutron star binary merger blast waves}
\shortauthors{Montes et al.}

\begin{document}

\title{Transport and Mixing of  \lowercase{{\it r}}-process elements in neutron star binary merger blast waves}
\author{Gabriela Montes\altaffilmark{1,2}, Enrico Ramirez-Ruiz\altaffilmark{1,2,3}, Jill Naiman\altaffilmark{2}, Sijing Shen\altaffilmark{4} and William H. Lee\altaffilmark{5}}

\altaffiltext{1}{Department of Astronomy and
  Astrophysics, University of California, Santa Cruz, CA
  95064}
\email{gmontes@ucsc.edu}
\altaffiltext{2}{Harvard-Smithsonian Center for Astrophysics, ITC, 60 Garden Street, Cambridge, MA 02138, USA}
\altaffiltext{3}{Radcliffe Fellow}
\altaffiltext{4}{Kavli Institute for Cosmology and Institute of Astronomy, University of Cambridge, Madingley Road, Cambridge CB3 0HA, UK}
\altaffiltext{5}{Instituto de Astronom\'ia, Universidad Nacional Aut{\'o}noma de M{\'e}xico, A. P. 70-264 04510 D. F. Mexico}

\begin{abstract} 
\noindent The $r$-process nuclei are robustly synthesized in the material ejected during a neutron star binary merger (NSBM), as tidal torques transport angular momentum and energy through the outer Lagrange point in the form of a vast  tidal tail. If NSBM are indeed solely responsible for the solar system $r$-process
abundances, a galaxy like our own would require to host a few NSBM  per million years, with each event ejecting, on average,
about $5 \times 10^{-2}M_\sun$ of $r$-process material.   Because the ejecta velocities in the tidal tail are significantly  larger than in ordinary supernovae, NSBM deposit  a comparable amount of energy  into the interstellar medium (ISM). In contrast to  extensive   efforts studying  spherical models for supernova remnant  evolution,  calculations quantifying the impact of NSBM ejecta  in the ISM have been lacking.  To better understand their  evolution in  a cosmological context, we perform a suite of three-dimensional hydrodynamic simulations with optically-thin radiative cooling  of isolated NSBM ejecta expanding in environments with conditions adopted  from  Milky Way-like galaxy simulations.  Although the  remnant  morphology is  highly complex at early times,
the subsequent radiative evolution that results  from thermal instability in atomic gas  is remarkably similar to that of a standard supernova  blast wave.  This implies that sub-resolution  supernova  feedback models can be used  in galaxy-scale simulations that  are unable to  resolve the key evolutionary phases of  NSBM  blast waves.  Among other quantities, we examine the radius, time, mass and  kinetic energy content  of the NSBM remnant  at shell formation as well as  the momentum injected to the ISM. We find that the shell formation epoch is attained when the swept-up mass is about $10^3\;M_\odot$, at this point the mass fraction of $r$-process material is drastically enhanced up to two orders of magnitude in relation  to a solar metallicity ISM. 
\end{abstract}


\section{Introduction}
The specific physical conditions and nuclear physics pathways  required for the $r$-process were originally identified in the pioneering work by \citet{burbidge1957}. But the particular astrophysical site remains open to more than one interpretation \citep{Sneden08}. Early  work on the subject identified  both type II supernovae \citep{woosley1994, takahashi1994} and neutron-star binary mergers \citep[NSBMs;][]{lattimer1977, freiburghaus1999} as likely candidate events to hold r-process. NSBMs are significantly  much rarer than type II supernovae \citep[e.g.][]{{argast2004}} and take place far from their birth places, reaching distance up to a few~Mpc from their host halo \citep[e.g.][]{kelley2010}. The two mechanisms  synthesize different quantities of $r$-process material, about $10^{-5}\,\rm{M_\odot}$ and  $10^{-2}\,\rm{M_\odot}$ for each type II supernovae and NSBM, respectively \citep[e.g.][]{cowan2004}. 
These differences should give rise to clear signatures  in the enrichment pattern of $r$-process elements in galaxies and  may ultimately help constrain the dominant production mechanism \citep{shen2014}.

With many difficulties getting  the necessary conditions to produce the $r$-process in  type II supernova winds \citep{qian1996, takahashi1994} ,
the NSBM  model has recently been extensively  studied and  shown to be a viable alternative.   The $r$-process nuclei have been found to be  robustly synthesized  \citep{metzger2010, roberts2011, korobkin2012, bauswein2013, grossman2014, ramirez-ruiz2015,goriely2015, mendoza2015}
and the  predicted galactic enrichment history of NSBMs  is consistent  with the abundance patterns observed in halo stars \citep{shen2014,  tsujimoto2014, vandevoort2015, vangioni2016}.
This suggests that the injection of $r$-process material by  NSBMs   has been operating  in a fairly robust manner  over long periods of time in galactic  history, while the resultant  chemical abundance dispersions in $r$-process elements such as Eu  suggests an early, chemically unmixed and inhomogeneous early Milky 
Way galaxy. At later times, these localized inhomogeneities would fade out as  more events happen and $r$-process products are given more time to be transported  and mixed  throughout the galaxy. 

An accurate treatment of the  evolution of the ejecta in NSBMs  is  thus crucial not only  for models of  electromagnetic transients \citep{nakar2011, rosswog2014, kelley2013, metzner2012} following coalescence but also for  models of $r$-process enrichment in galaxies.  In contrast to the extensive efforts developing models for supernova  remnant evolution, \citep{cox1974, mckee1977,blondin1998, joung2006,martizzi2015,kim2015} studies quantifying the impact of NSBM remnants   in a cosmological context  have not been carried out. The efficacy of transport and mixing of $r$-process in NSBM remnants has not been properly  quantified  mainly due to the highly inhomogeneous initial conditions  and  the  excessive radiative losses expected during the shock propagation.  

In this paper, we present the results of a series of controlled three-dimensional hydrodynamic simulations of NSBM remnants. 
To calculate the ejected mass and initial structure of the tidal tails, we make use of three-dimensional smoothed particle hydrodynamics simulations  \citep[][]{roberts2011} of  NSBMs (Section~\ref{mergers}).
 The resultant  homologous structure of the tidal tail is then mapped into an adaptive mesh refinement (AMR) simulation with optically-thin radiative cooling (Section \ref{hydro}).   Our goal is to understand  the evolution  of  single NSBM remnants,  which is quantified   in  Section~\ref{dynamics},  and how it might  depend on  the properties of the surrounding medium,  which are derived using   a cosmological simulation of the formation of the Milky Way (Section~\ref{amb}).  Our simulations of isolated NSBMs are used to construct  sub-resolution
prescriptions for galaxy-scale simulations with inadequate resolution
to properly define the the cooling radius of NSBM blast waves. Discussion of the results as well as detail comparisons with studies  from spherical models    are presented in Section~\ref{dis}.

\section{Methods}

\subsection{Initial Conditions}\label{mergers}

Tidal tails are a common feature formed during
mergers and collisions between compact objects. These are typically a few thousand
kilometers in size by the end of the merger event in the case of neutron star  disruptions by black holes  and NSBMs \citep{lee2007, faber2012}.
Some  small fraction of the material ($10^{-3} -10^{-2}M_\odot$) in the tails
 is actually unbound and will escape to the surrounding medium. The  exact mass and  structure (density and velocity distribution) of the ejected material depends on the  mass ratio and  details of the equation of state \citep[EoS; e.g.][]{roberts2011}

\begin{figure}[t!]
  \centering
    \includegraphics[width=0.5\textwidth]{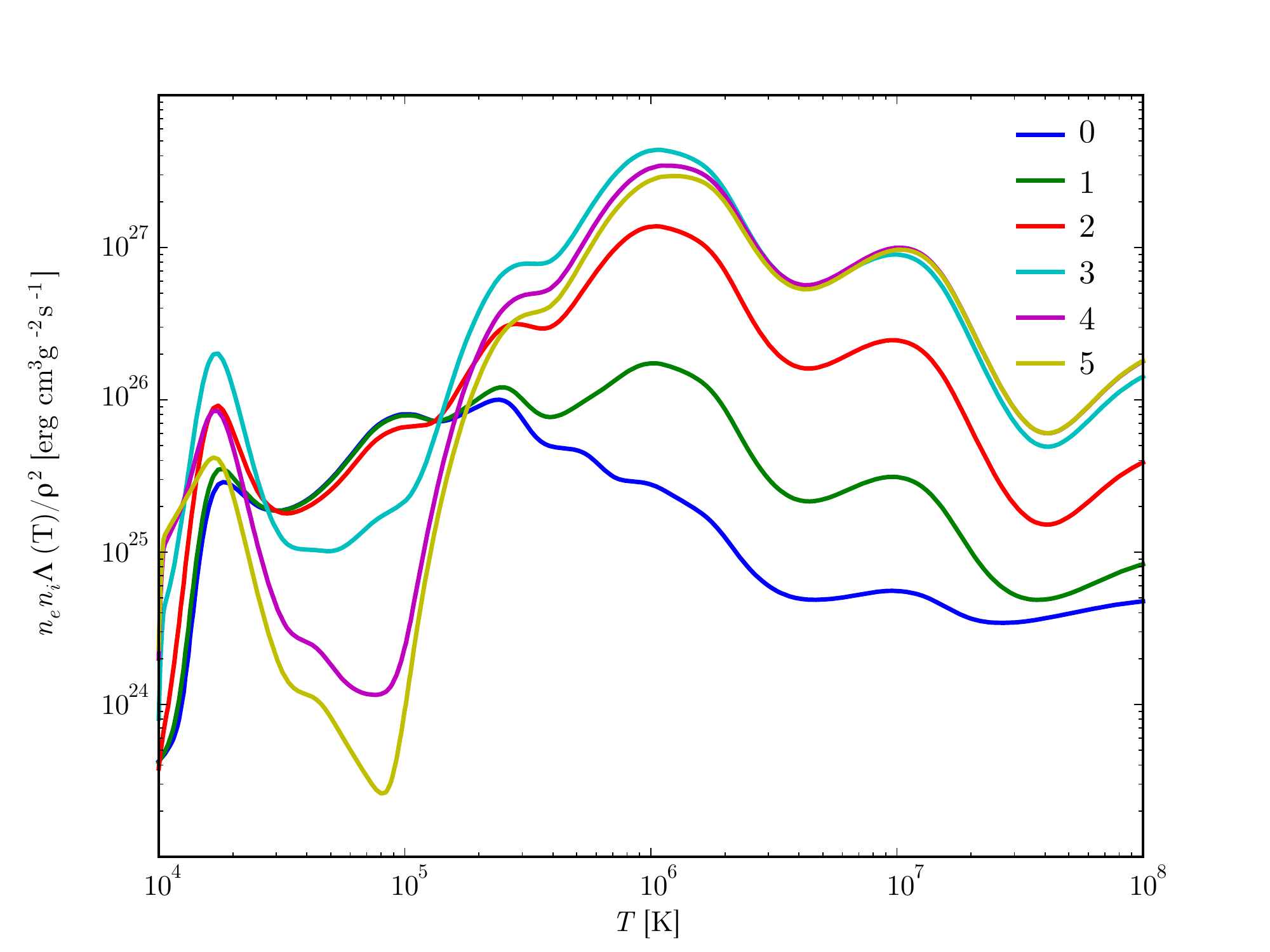}
 \caption{Cooling functions constructed using  the ion-by-ion cooling efficiencies for low-density gas derived by \citet{gnat2012} and assuming that the full  radiative spectra  of $r$-process material  is similar to that of Fe.  The computed cooling  curves with enhanced metal  abundances are plotted  here as a function of the augmented Fe mass fraction  $\xi_{\rm metal}=(\chi_{\rm Fe} / \chi_{\rm Fe, \odot})=10^0,10^1,10^2,10^3,10^4,10^5$. The standard case of solar abundances is denoted here as $\log (\chi_{\rm Fe} / \chi_{\rm Fe, \odot})=0$. Once the cooling function is tabulated, the temperature variation within the expanding, cooling shocked  layer in the NSBM remnant can be calculated.}	
 \vspace{0.3cm}
 \label{fig:coolingcurves}
\end{figure}

To calculate the  density and velocity structure of the unbound tidal tails we use a similar method to the one described by \citet{roberts2011}.  
We use three-dimensional smoothed particle hydrodynamics (SPH) simulations \citep{lee2007, lee2010} to follow the  merger of two neutron stars during which the geometry, densities, and timescales change violently \citep{rasio1994, lee2000, rosswog2003}.  As a representative  example, we study the dynamics of the $M_{\rm r}=0.05 M_\odot$ material ejected during a NSBM with mass ratio $q = M_2 /M_1 =0.88$ \citep{roberts2011}. A hybrid EoS, similar to that implemented  by \citet{shibata2005}, is used which combines the cold Friedman-Pandharipande-Skyrme nuclear EoS with an ideal gas component. Once the initial dynamical interaction is realized, the fluid elements in the  unbound tails are verging on
ballistic trajectories, moving primarily under the influence of the central mass potential.   The hydrodynamical calculations are stopped only after the expansion of the unbound material becomes homologous. 
Figure 1 in \citet{roberts2011} depicts the structure of the tidal tails at homology for a wide range of NSBMs. A clear progression is observed from equal mass tails formed in the $q = 1.0$ case to almost no secondary tail produced in the $q = 0.88$ case studied  here.

\subsection{Hydrodynamical  Evolution of NSBM Remnants}\label{hydro}

Here we follow the expansion of the tidal tail produced by  a NSBM with $q = 0.88$  in three dimensions with the parallel, adaptive-mesh, hydrodynamical code FLASH \citep{fryxell2000}.  
We evolve the ideal  fluid  in three dimensional cartesian coordinates using a metallicity-dependent cooling function, which is  constructed using 
the ion-by-ion cooling efficiencies for low-density gas derived by \citet{gnat2012} for gas temperatures between $10^4$ and $10^8$ K. As it is generally adopted, we do not include metal fine structure transitions or molecular line cooling and so the cooling function is effectively truncated below $T\approx  10^4$ K.

Figure \ref{fig:coolingcurves} shows
the cooling function for an optically thin thermal plasma with solar abundances assuming collisional equilibrium ({\it blue} curve).
The full radiative spectra of $r$-process  material at the relevant densities and temperatures is not well known as complete atomic line lists for these heavy species are not available \citep{kasen2013, tanaka2014, lippuner2015}. 
We therefore simply assume that the energy integral of the full radiative spectra  of $r$-process material  is similar to that of Fe. The computed cooling  curves with enhanced Fe abundances are  plotted in Figure \ref{fig:coolingcurves}.  This assumption is, however, of no consequence at all as  radiative losses begin to influence the NSBM remnant  evolution only when the metal content is almost exclusively  dominated by the swept up gas (see Section \ref{dynamics} for specifics). As such,  the resulting evolution  is quantitatively 
similar to that computed using solar abundances, but note that we have implemented metal enhanced  cooling rates  here for completeness.

\begin{figure}[t!]
  \centering
    \includegraphics[width=0.38\textwidth]{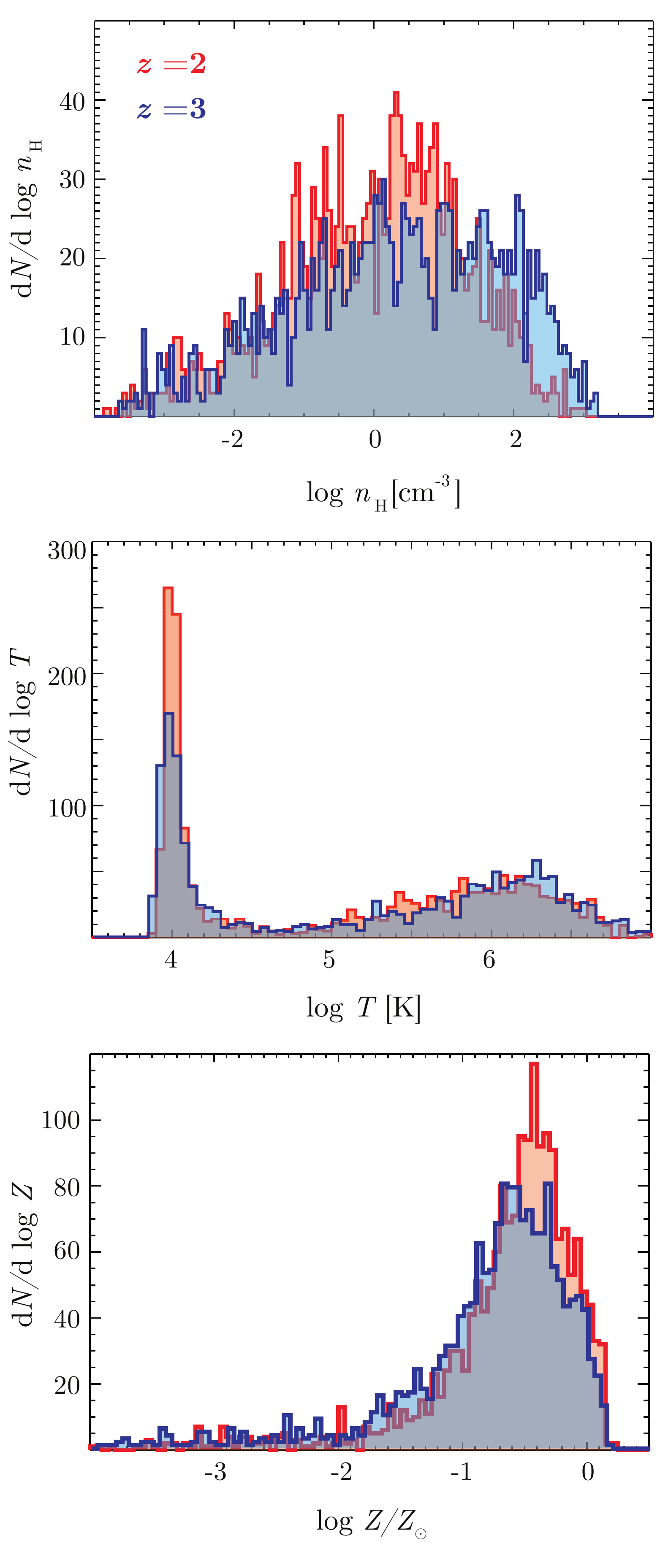}
 \caption{The gas density, temperature and metallicity  at the NSBM injection
sites in the Galaxy at redshifts $z=$ 2 and 3, as derived from the Eris simulation \citep{shen2014, guedes2011}. 
 The spatial distribution of 
NSBMs is assumed to broadly follow the stellar distribution of  the host
galaxy  while the injection rate is derived using the Eris star formation rate convolved  with a power-law delay-time distribution. Star We note that star formation in the simulation  occurs when cold ($T < 3\times 10^4$K)  gas  reaches  the  threshold  density and as such cooling is effectively truncated. }	
 \label{fig:env}
\end{figure}

\subsection{Properties of the Ambient  Medium}\label{amb}

\begin{table}[t]
  \begin{center}
\caption{Initial Conditions and Ambient Medium Properties}	
  \label{ta}
    \begin{tabular}{ l| l }
    \hline
    \hline
Tidal Tail & NSBM with $q=0.88$\\
\hline
    $M_{\rm r}$            & $5.8\times10^{-2}$~M$_\odot$ \\
    $E_{\rm k}$          &   $3.58\times10^{51}$~erg  \\
\hline    
ISM\\
\hline
    $n_{\rm H}$          & 1$\;\rm{cm^{-3}}$   \\
    $T$      & $10^4$~K \\  
    $Z    $    &  $Z_\odot$  \\ 
\hline
Rarefied\\
\hline
    $n_{\rm H}$          & $10^{-4}\;\rm{cm^{-3}}$   \\
    $T$      & $10^6$~K \\     
    $ Z$         & $Z_\odot$  \\
\hline
Dense \\
\hline
    $n_{\rm H}$          & $10^{2}\;\rm{cm^{-3}}$   \\
    $T$      & $10^4$~K \\      
    $ Z$         & $Z_\odot$  \\
\hline
    \end{tabular} 
  \end{center}

\end{table}

The evolution of a NSBM remnant  depends on the character  of
the ambient  medium. In Figure \ref{fig:env} we show the expected properties of the  gas in and around  the  NSBM injection sites, derived using the  cosmological zoom-in simulation Eris,  which at redshift $z = 0$ is a close analog of the Milky Way \citep{guedes2011}. The merger rate is inferred  by convolving the star formation history with a standard  delay-time distribution of mergers modeled by a power-law \citep{piran1992, kalogera2001, shen2014}.  The resulting NSBM  history in the simulation is
shown in Figure 1 of  \citet{shen2014}.  The spatial distribution of NSBM
mergers is  then assumed to  roughly follow the stellar distribution.  Using these two key model ingredients, the number and location of NSBMs
 can then be  estimated, which in turn can be used to infer the  density, temperature and metallicity  of the surrounding  ambient gas (Figure~\ref{fig:env}). 
Motivated by this, we run three different types of simulations: 
\begin{itemize}
\item isolated NSBM in a homogeneous ISM with $n_{\rm H}=1\;\rm{cm^{-3}}$, $T=10^4$~K and $Z=Z_\odot$;
\item  isolated NSBM in a homogeneous, dense   ISM with  $n_{\rm H}=100 \;\rm{cm^{-3}}$,  $T=10^4$~K and $Z=Z_\odot$;
\item isolated NSBM in a homogeneous, rarefied  ISM  with  $n_{\rm H}=10^{-4}\;\rm{cm^{-3}}$, $T=10^6$~K and $Z=Z_\odot$.
\end{itemize}
The parameters of the different simulations are outlined in
Table~\ref{ta}. For comparison, standard blast wave simulations initiated by injecting the same  total energy in a spherical region are computed  in order to determine how the
evolution of a NSBM is  initially altered by its non-uniform original structure. In all cases, we consider computational boxes filled
with an  ISM initially in pressure equilibrium and adopt
a refinement scheme based on pressure and density  gradients which refines
around the expanding shock.

\section{Hydrodynamical Evolution of NSBM Remnants}\label{dynamics}
The evolution of an isolated NSBM remnant expanding  into  a uniform
medium can be broadly characterized by the well-known evolutionary stages of a supernova remnant (SNR):
\begin{itemize}
\item the free expansion phase, during which the mass of the
tidal ejecta is larger than the mass of the swept up ISM; 
\item the energy conserving phase, during which radiative losses
are not important; 
\item the  cooling-modified pressure-driven snowplow phase, during which  shell formation occurs;
\item and  the  final momentum-conserving expansion phase.
\end{itemize}
\begin{figure}
\centering
                   \includegraphics[width=0.36\textwidth]{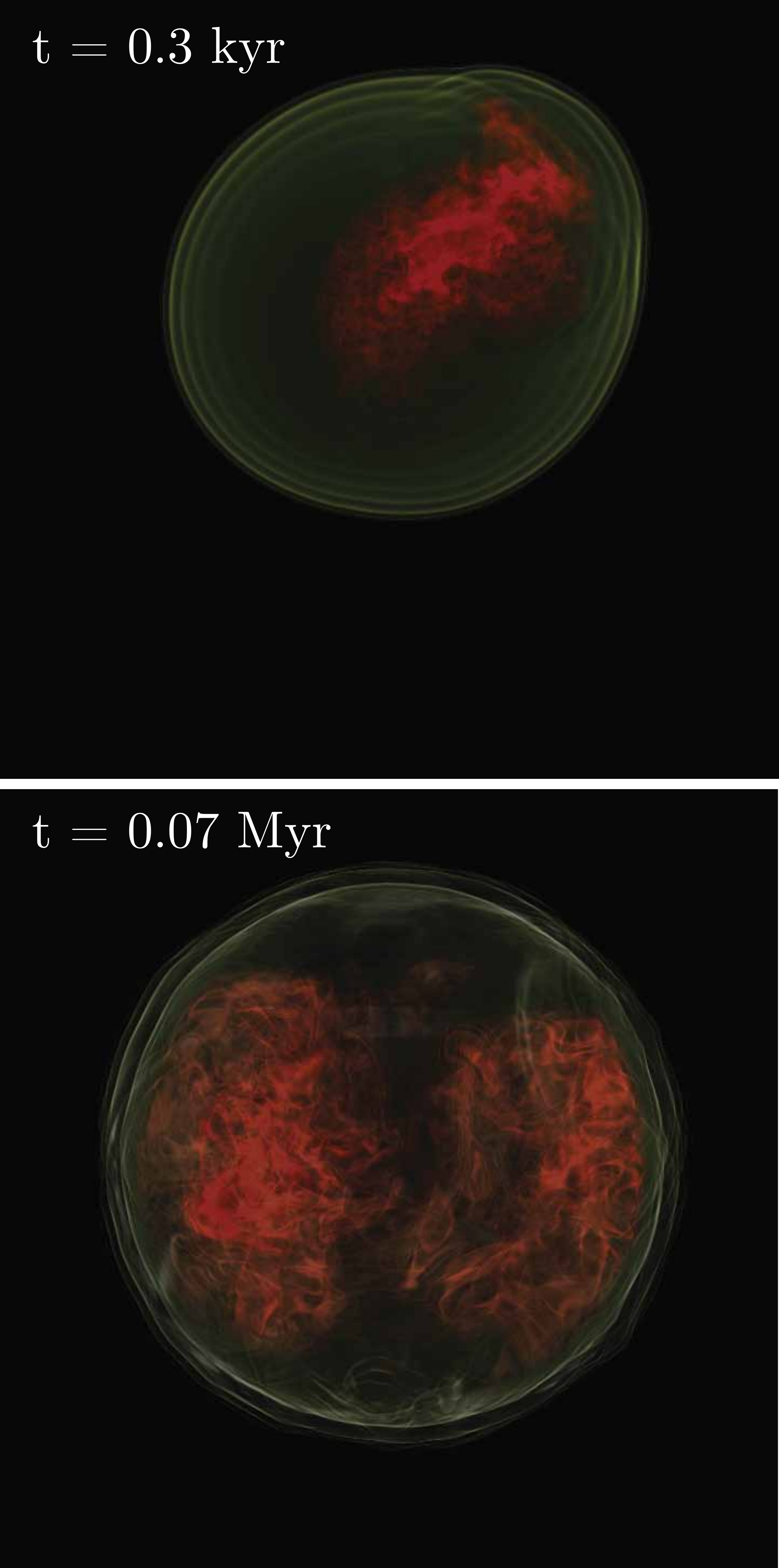}
        \caption{The evolution of a NSBM remnant expanding into a uniform ISM (fiducial run labelled {\it ISM} in Table~\ref{ta}). {\it Top panel}: Density contours  (green) are plotted for 12 levels from $\log(\rho) = -23.4$ to $-20.8$ g cm${-3}$. The mass fraction of metals in the ejecta, are denoted as $\chi_{\rm metal}$ and is assumed to have the cooling properties of Fe. The mass fraction of metals in the ejecta, denoted by $\chi_{\rm metal}$, is represented by surfaces (red)  plotted with 16 levels from $\log(\chi_{\rm metal}) = -6$ to $0$. The box size is 2.4~pc. {\it Bottom panel}:  Density surfaces  (green) are plotted for 12 levels from $\log(\rho) = -26$ to $-20.7$ g cm${-3}$.   $\chi_{\rm metal}$ contours  (red) are plotted with 16 levels from $\log(\chi_{\rm metal}) = -6$ to $0$. The box size is 20~pc. Total box size is 40~pc with  $512\times512$ cells on the coarsest grid and 11 initial levels of refinement corresponding to a maximum resolution of $1.504\times10^{16}$~cm. Levels of refinement are decreasing to reach a maximum level of eight with a maximum resolution of $1.203\times10^{17}$~cm for the bottom snapshot.}\label{fig:3d}
\end{figure}
However, differences in ejected mass and  initial structure  modifies the evolution of a NSBM remnant relative to that  of a SNR, as 
illustrated in Figure~\ref{fig:3d} for our fiducial run (labelled {\it ISM} in Table~\ref{ta}).  The top panel shows the  evolution at time $t = 0.3$ kyr,  long after the free expansion phase and during the energy conserving phase. The kinetic energy of the NSBM $q=0.88$ tidal tail is $3.58 \times 10^{51}$ erg. The ensuing strong blast wave converts much of this
energy into thermal energy, which leads to lateral expansion\footnote{Similar to the case of remnants arising from  jet-driven supernova explosions \citep{ramirez-ruiz2010,gonzalez2014} or from  the unbound debris of stars  disrupted by massive black holes \citep{kasen2010,guillochon2015}.}.  As the NSBM remnant  ages, the thermal energy is lost
to radiative cooling, the shock slows down, and shell formation is established. The bottom  panel in Figure~\ref{fig:3d} shows the effect of the blast wave
propagation at time $t = 0.07$ Myr, before  the onset of the remnant's  cooling-modified snowplow phase.

\begin{figure}[t!]
\centering
                    \includegraphics[width=0.4\textwidth]{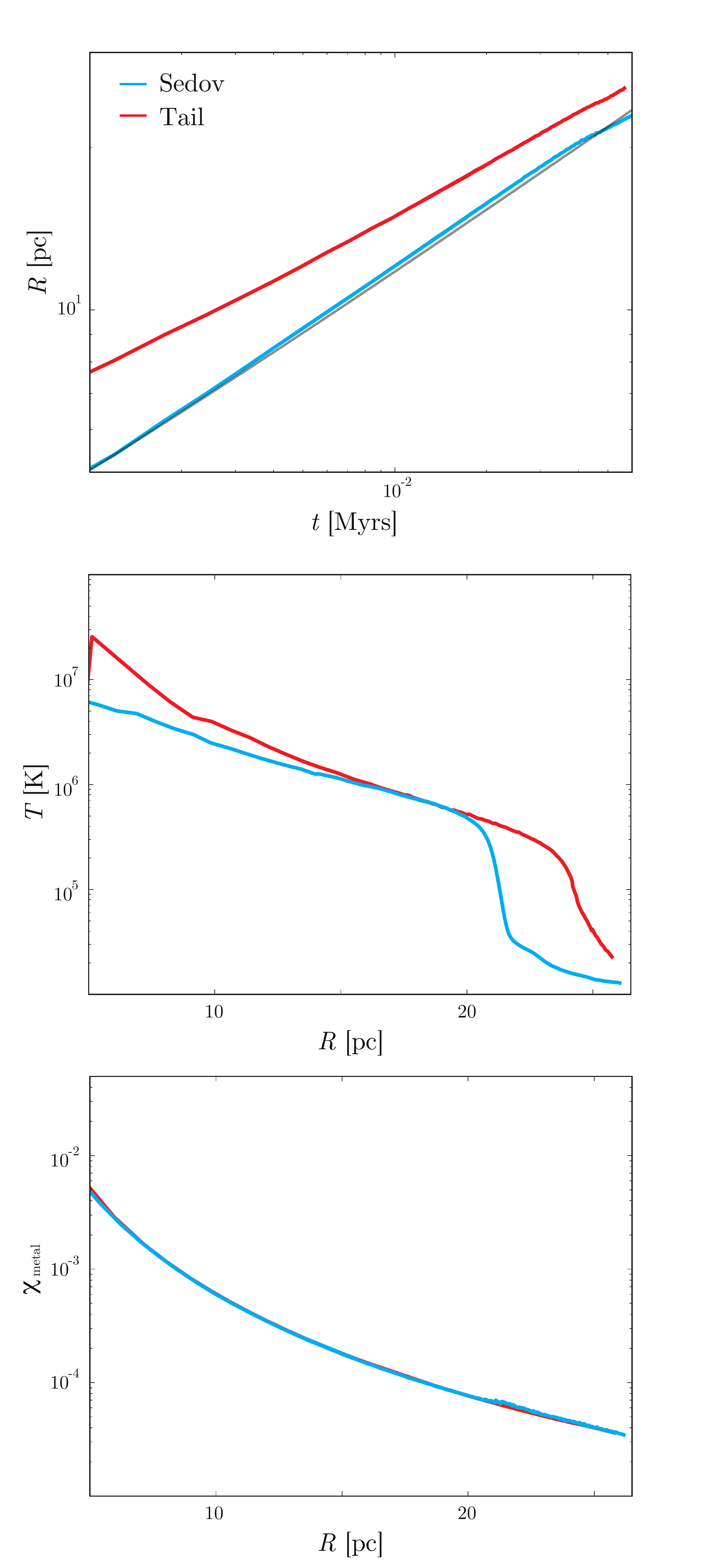}
        \caption{Evolution of the forward shock  $R$ with time $t$ for the NSBM remnant case depicted in Figure~\ref{fig:3d} (labelled as {\it Tail}). The evolution of a spherical blast wave with identical energy and mass content, labelled as {\it Sedov}, is  plotted for comparison. The analytical Sedov-Taylor solution $R(t)$ is shown as a grey line. Also shown are the temperature $T$ and metal mass fraction $\chi_{\rm metal}$ with $R$. The NSBM remnant  propagates faster owing to the reduced rate of  ambient mass sweeping, which is subsequently increased as the remnant becomes  progressively more spherical. As a result, shell formation is delayed relative to the spherical case.}
        \label{fig:ism}
\end{figure}

The top panel of Figure~\ref{fig:ism} shows the evolution of the forward shock radius $R$ with
 time $t$  for the NSBM remnant  case shown in Figure~\ref{fig:3d}. For comparison, the evolution 
of a spherical blast wave with the same initial mass and energy is also shown together with  the analytical Sedov-Taylor solution.  For spherical simulations, $R$ is identified by measuring
spherically averaged  profiles, while for the NSBM simulations we approximate $R$ as the radius
of the sphere enclosing 90\% of the total energy (no significant difference in the measured radio was obtained from slightly different percentages). The radial  temperature $T$ and metal mass fraction $\chi_{\rm metal}$ profiles are then calculated  by measuring
the spherically averaged values of such quantities at $R$.

\begin{figure}[t!]
\centering
                   \includegraphics[width=0.4\textwidth]{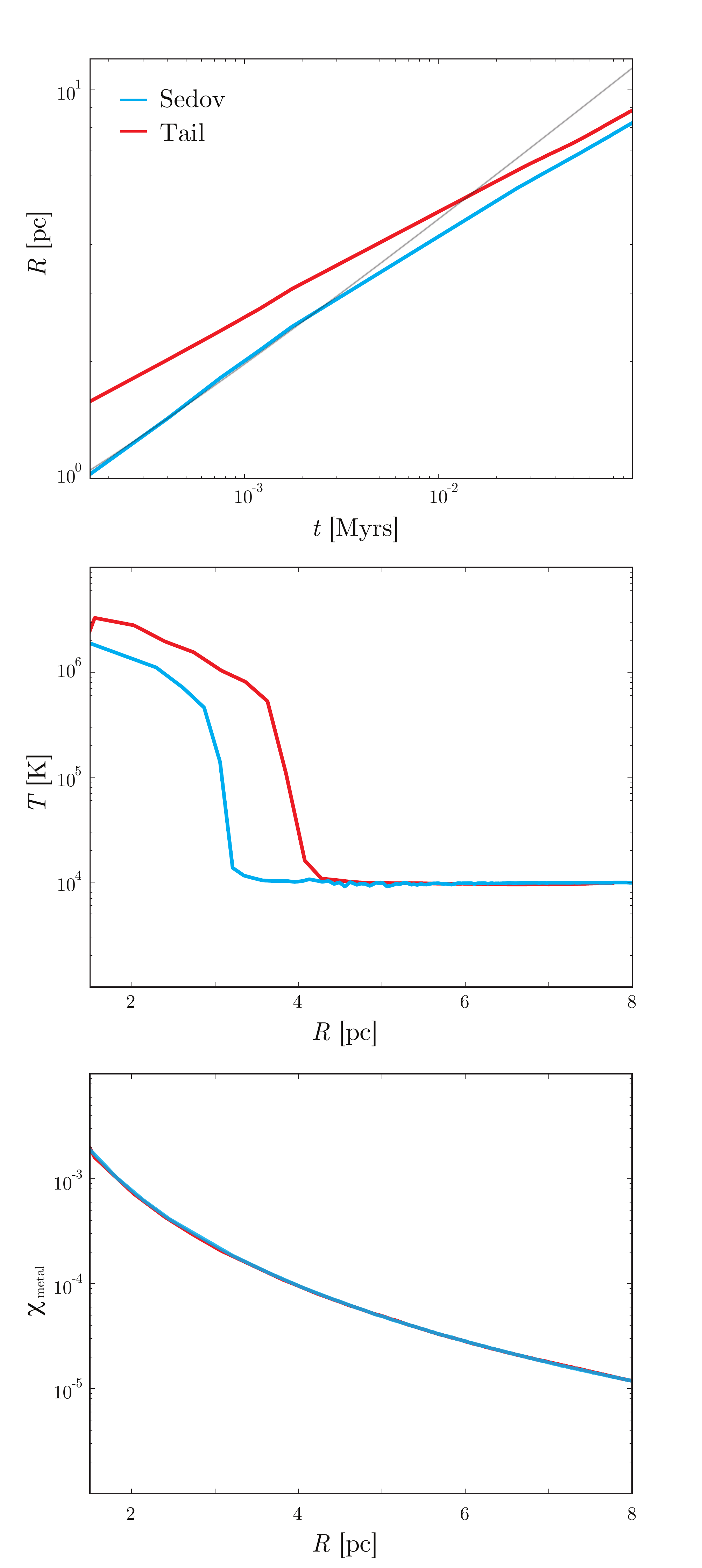}
        \caption{Similar to Figure~\ref{fig:ism} but for expansion into a dense ambient medium (labelled {\it Dense} in Table~\ref{ta}).}\label{fig:dense}
\end{figure}

\begin{figure}
\centering
                   \includegraphics[width=0.4\textwidth]{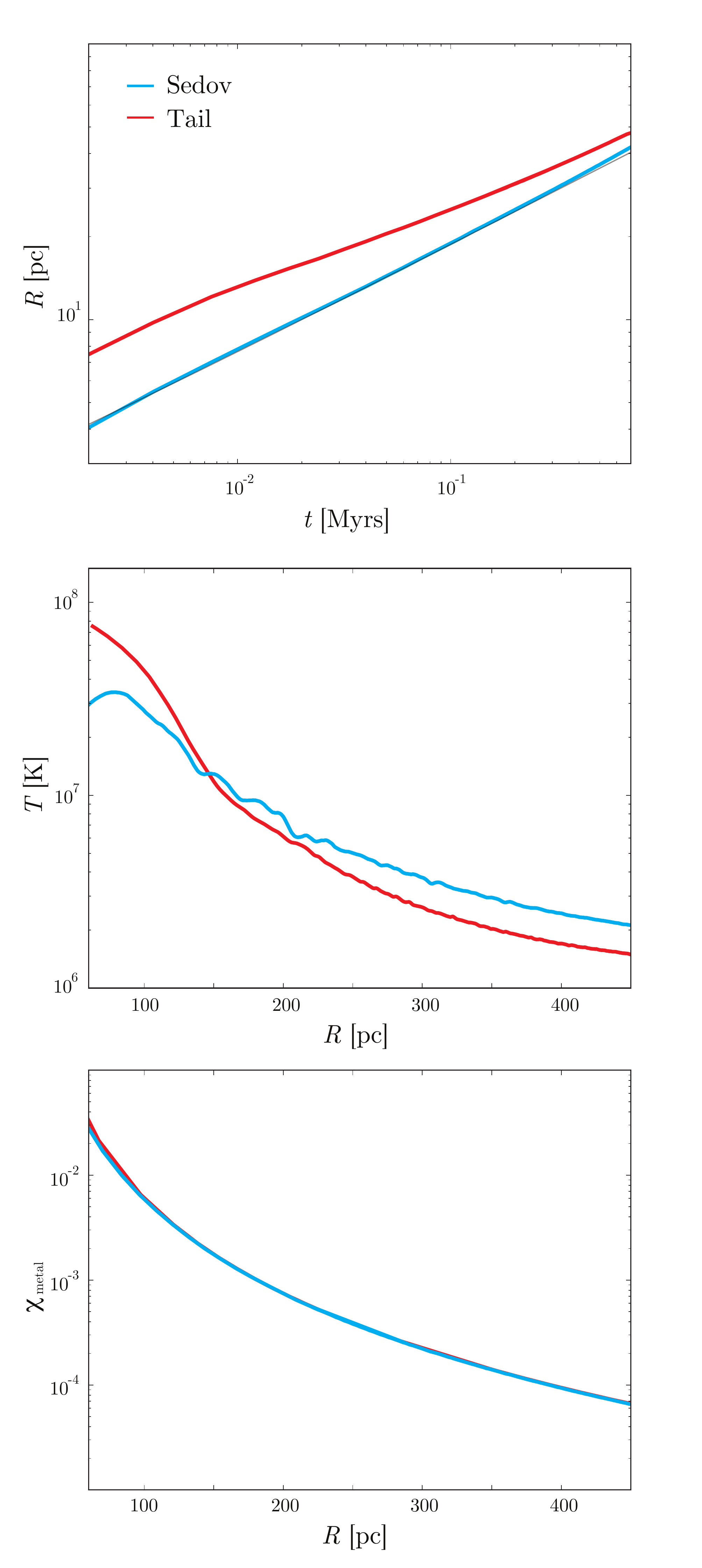}
        \caption{Similar to Figure~\ref{fig:ism} but for expansion into a rarefied and hot ambient medium (labelled {\it Rarefied} in Table~\ref{ta}).}\label{fig:rarified}
\end{figure}

The initial configuration of the ejected tidal tail material  is highly  inhomogeneous, as can be seen  in Figure~\ref{fig:3d}, with shock expansion velocities being larger along the original orbital plane  of the NS binary. As a result, the rate of ambient  material  swept up by the blast wave is reduced relative to the spherical case until  lateral expansion  becomes important.  The Sedov-Taylor solution is an attractor and the NSBM remnant evolution slowly  adjust
to match this spherical,  energy conserving solution. As shown in Figure~\ref{fig:ism}, the NSBM remnant becomes radiative  before matching   the Sedov-Taylor solution, which  increases  the local cooling time  and leads to
a decrease in  radiative losses when compared to the spherical case. Similar behavior is also seen in simulations 
 of a NSBM remnant expanding into a dense ambient medium (Figure~\ref{fig:dense}) and into a rarefied, hot  external environment (Figure~\ref{fig:rarified}).

In a uniform medium, the shell formation epoch, which occurs when a SNR  becomes radiative,  is usually well characterized  by the time  the mass of swept up material  attains $M_{\rm s}\approx  10^3 M_\odot$.  Previous studies  of SNRs expanding in a homogeneous  ambient medium \citep{cioffi1988, thornton1998} approximate  the cooling radius as 
\begin{equation}
R_{\rm s}\approx 14 \left({E_{\rm tot} \over 10^{51}\;{\rm erg}}\right)^{2/7} \left({n_{\rm H} \over 1\;{\rm cm}^{-3}}\right)^{-3/7} \left({Z \over Z_\odot}\right)^{-1/7}\;{\rm pc}. 
\end{equation}
Here we evaluate the  radius $R_{\rm s}$, swept up mass $M_{\rm s}$, and kinetic energy $E_{\rm k}$ of the NSBM remnant at shell formation. The results for all simulations  are given in
Table~\ref{table:rs}.  The radius, total remnant mass, and outward radial momentum at shell formation are similar  to those obtain using spherical  simulations and as such are close to the analytic estimates. Despite the initial differences in ejecta mass and geometry, our conclusions regarding shell formation and momentum injection in NSBM remnants  are quite similar to those obtain for SNRs.
\begin{table*}
  \begin{center}
   \caption{properties of the blast wave at  shell formation}
    \begin{tabular}{ l| ccccc }
    \hline
 \hline
Simulation & $R_{\rm s}/\rm{pc}$  & $M_{\rm s}/\rm{M_\odot}$ & $E_{\rm k,tail}/E_{\rm k,sedov}$ &< $\chi_{\rm r}/\chi_{\rm r,\odot}$> & <$\chi_{\rm metal}$>\\
\hline
  ISM         & 26.5                           & $1.93\times10^3$        &   1.04                                 &   140.87                                   &$3.32\times10^{-5}$\\
  Rarefied   &   471.72                     & $1.09\times10^3$        &    1.04                                 &   243.90                                   &$5.73\times10^{-5}$\\
  Dense & 5.7                                 & $1.92\times10^3$        &    1.06                                 &  139.02                                    &$3.27\times10^{-5}$\\
  
\hline

    \end{tabular} 
  \end{center}
  \label{table:rs} 
\end{table*}


At the time of shell formation, the mass fraction $\chi_{\rm metal}$ arising from metals ejected during the NSBM is about $5\times 10^{-5}$. We thus only expect metal cooling from $r$-process material to influence the dynamics  of NSBM remnants if they expand into an ambient medium with  $Z \lesssim 10^{-3} Z_\odot$, under the assumption  that the metal cooling function   of $r$-process material  is similar to that of Fe (Section~\ref{hydro}).  The $r$-process enrichment of the  gas depends on how efficiently the metals are mixed with the ambient material swept up by the blast wave. At the time of shell formation, we find the mass fraction of $r$-process material  in the NSBM remnant to be  drastically enhanced  in relation to solar metallicity  to about  $10^2\,(Z/Z_\odot)$. 

\section{Discussion and Conclusions}\label{dis}
The importance of SNR evolution  for detailed models of the ISM  but also for models of galaxy evolution  and chemical enrichment has been appreciated for decades. In recent years, it has become increasingly clear that an accurate treatment of NSBM remnant evolution is crucial not just for detailed models of electromagnetic transients \citep{nakar2011, rosswog2014, kelley2013, metzner2012}  but also for models of heavy element enrichment \citep{shen2014,  tsujimoto2014, vandevoort2015}. NSBMs are thought to play a predominant role in creating \citep{metzger2010, roberts2011, korobkin2012, bauswein2013, grossman2014, ramirez-ruiz2015,goriely2015, vangioni2016} and dispersing $r$-process elements, yet modern cosmological simulations are still  not able to resolve  the expansion of SNRs. Motivated by this, we  have perform a series of  three-dimensional hydrodynamic simulations of isolated NSBM remnants expanding in  environments with thermodynamical properties  similar to those found in cosmological simulations (Section~\ref{amb}).   Particularly important is the use of realistic merger simulations \citep{roberts2011} for constructing the  initial conditions characterizing the structure of the ejected tidal debris (Section~\ref{hydro}).  We have centered our attention  in particular on the evolution  of NSBM
remnants during the well-known stages  that are commonly used to describe the evolution of SNRs.  

\begin{itemize}
\item First, a free expansion phase is observed, during which the mass of the
ejecta is larger than the mass of the swept up gas. This evolutionary phase is significantly shorter than that of a typical SNR due mainly  to the low mass content of the  ejecta ($5.8\times10^{-2}~M_\odot$) and, to a lesser degree,  the initial complex geometry. 

\item Secondly, an energy conserving phase takes place, during which radiative losses
are negligible (Figure~\ref{fig:3d}). During this phase the rate of mass swept up by the blast wave is reduced when compared to a standard SNR but is then subsequently  increased as the
 NSBM remnant becomes progressively more spherical.  However, radiative losses  begin to influence the  NSBM remnant evolution before it is accurately described  by a standard spherical solution (Figure~\ref{fig:ism}).  In addition, the overall evolution takes
slightly longer, with the cooling time reaching its  critical value only when the shock has travelled a distance that is a  few times larger than in the spherical case. This is observed in all simulations listed in Table~\ref{ta}.  
 
 \item Thirdly, a pressure-driven snowplow phase sets in,  during which
radiative losses begin to influence the evolution of the NSBM remnant 
and  shell formation is found to set in at a time when the swept up mass attains a value of $M_{\rm s}\approx 10^3 M_\odot$, as it is commonly found in SNRs (Table~\ref{table:rs}).  The radius, total  swept-up mass, kinetic energy and outward radial momentum at shell formation for NSBM remnants, we conclude,  are  close to the standard  estimates given for SNRs  \citep{cox1974, mckee1977,blondin1998, joung2006}. 

\item Finally, during the momentum conserving
phase, the evolution of a NSBM remnant  is remarkably  similar to that of a  SNR. This implies that sub-resolution supernova feedback models \citep[e.g.][]{martizzi2015,kim2015} can be accurately used in galaxy-scale simulations that are unable to resolve the early  evolutionary stages  of  a NSBM remnant.
 \end{itemize}
 
While our simulations of NSBM remnants have confirmed the similarities and highlighted  the differences of the well-known evolutionary stages of SNRs, 
one of the key distinct processes affecting their structure  is the contribution of $r$-process material to the cooling of the swept-up material.   As complete atomic line lists for these heavy species are not available \citep{kasen2013}, 
we have thus simply assumed that the energy integral of the full radiative spectra of $r$-process material is similar to that of Fe. Under these conditions, we find that  metal cooling from $r$-process material is not expected  to influence the dynamics  of NSBM remnants   expanding into an ambient medium with  $Z \gtrsim 10^{-3} Z_\odot$.  The resulting NSBM remnant evolution in these environments  should solely be determined by the metal cooling of the swept-up material. Yet,  the mass fraction of $r$-process material  at shell formation  within the remnant is expected to be severely enhanced  to about $2.5 \times 10^4\; (Z/ 10 ^{-2} Z_\odot)$, assuming a solar  abundance ratio.  In contrast,  with a  total $r$-process mass per supernova of $M_{\rm p}\approx  7.4 \times10^{-5} M_\odot$ \citep{shen2014}, we expect a local enhancement of only about $31.9 \; (Z/ 10 ^{-2} Z_\odot)$ when a  Type II  SNR attains shell formation.
 
Of particular relevance  in this case are the heavy element composition of  galactic halo stars with  metallicity regime $10^{-3} Z_\odot \lesssim Z \lesssim 10 ^{-2} Z_\odot$, which have been found to have  pure $r$-process products with a distribution that is characteristic of solar system matter but with a large star-to-star scatter in their  $r$-process  concentrations \citep{Sneden08}.  In  some stars with $Z\approx 10 ^{-2} Z_\odot$, the $r$-process enrichment  can be as large as $10^2$, assuming a solar  abundance ratio.  The presence of $r$-process material in these stars with a distribution that is characteristic of solar abundance ratios  illustrates  that $r$-process enrichment  has operated in a fairly robust manner, while their abundance dispersions  further  suggests  that $r$-process production sites must be rare and locally very enhanced, as expected from NSBM enrichment \citep[e.g.][]{shen2014}.  These localized inhomogeneities would then be smoothed out as  more NSBMs take place and $r$-process material is  given more time to be transported and mixed  throughout the early Milky Way. Metal feedback from NSBM remnants  is thus essential to understand  the formation of $r$-process enhanced stars  in  galaxies. Despite the initial differences, our conclusions regarding the efficacy of the  transport and mixing of $r$-process material  by NSBM remnants  in a cosmological context should be  similar to that expected from single SNRs.

\acknowledgments 
We thank R. Cooke, D. Kasen,  E. Kirby and  L. Roberts  for insightful
discussions and acknowledge financial support
from the David and Lucile Packard Foundation, NSF
(AST0847563) and UCMEXUS (CN-12-578).  G. MONTES acknowledges to AAUW American Fellowship 2014-15. We gratefully acknowledge the hospitality of the Aspen Center for Physics, the Radcliffe Institute for Advanced Study and  the  the DARK
Cosmology Center while completing this work.

\end{document}